# The dynamical-quantization approach to open quantum systems

A. O. Bolivar[a)]

Departamento de Física, Universidade Federal de Minas Gerais, Caixa Postal 702, 30123-970, Belo Horizonte, Minas Gerais, Brazil

## Abstract

On the basis of the dynamical-quantization approach to open quantum systems, we can derive a non-Markovian Caldeira—Leggett quantum master equation as well as a non-Markovian quantum Smoluchowski equation in phase space. On the one hand, we solve our Caldeira—Leggett equation for the case of a quantum Brownian particle in a gravitational field. On the other hand, we solve our quantum Smoluchowski equation for a harmonic oscillator. In both physical situations we come up with the existence of a non-equilibrium thermal quantum force. Further, as a physical application of our quantum Smoluchowski equation we take up the phenomenon of escape rate of a non-inertial Brownian particle over a potential barrier.

a) *E-mail address* : bolivar@cbpf.br or bolivar@fisica.ufmg.br (A. O. Bolivar)

## 1. Introduction

From the mathematical point of view, the dynamics of an isolated material point is deterministically described by the Schrödinger equation

$$\frac{d\Psi(t)}{dt} = \mathcal{H}\Psi(t),$$

where the Hamiltonian operator $\mathcal{H}$ acting upon the time-dependent function $\Psi(t)$ can carry features inherent in the particle such as its mass $m$, its position $x$ or its momentum $p$. Moreover, the mathematical structure of $\mathcal{H}$ also relies on the phenomenological parameter ħ, dubbed the Planck constant $h$ divided by $2\pi$, which in turn is responsible for the signature of the quantum world. Most especially, the Schrödinger functions $\Psi(t)$ account for the remarkable phenomenon of superposition or interference of quantum states which is in the core of current researches on quantum computation [1].

Yet from the physical standpoint, a quantum system cannot be imagined as being in isolation from its surroundings. In truth, it only comes into being as far as its interaction with a certain environment (e.g., a measuring apparatus) is concerned [2,3]. Accordingly, a quantum system must be actually idealized as an open (non-Hamiltonian) quantum system comprising of a tagged particle immersed in a generic quantum environment. The jittering movement undergone by such a tagged particle is dubbed quantum Brownian motion which is to be mathematically described by quantum master equations of the general form

$$\frac{d\rho(t)}{dt} = \mathcal{L}\rho(t),$$

where the non-Hamiltonian superoperator $\mathcal{L}$ (the so-termed Liouvillian of the quantum open system) acting on the von Neumann density operator $\rho(t)$ bears some environmental features, such as coupling constants (a sort of friction constant) and the fluctuation energy accounting for the existence of the quantum Brownian movement, as well as some properties inherent in the tagged particle such as its mass $m$, its position $x$, and the Planck constant ħ. In theory of quantum open systems the pivotal issue is therefore the following [4]: *How can we build up or derive some physically meaningful non-Hamiltonian Liouville operators* $\mathcal{L}$*?*

From the mathematical viewpoint, the superoperator $\mathcal{L}$ can be algebraically built up as Lindblad operators by beginning with a system-environment model Hamiltonian, and then making the Born and Markov approximations. This approach describes a sort of Markovian interaction between the non-Hamiltonian system (e.g., the Brownian particle) and its environment. In addition, it is assumed that the system and environment begin in a product state, i.e., they are initially independent and non-interacting [2—6]. Applications of the Lindblad formalism can be found, for instance, in quantum optics in which the environment is represented by a quantized radiation field while the Brownian particle is deemed to be an atom or a molecule. The corresponding master equation is called quantum optical master equation [6,7].

Nevertheless, it has been argued that "*one should not attribute fundamental significance to the Lindblad master equation*" because "*the Lindblad theory is not applicable in most problems of solid state physics at low temperatures for which neither the Born approximation is valid nor the Markov assumption holds*" [8]. Furthermore, recent controversies [9] seem to point out the inadequacy of the Lindblad approach to fathoming the true physics of open quantum systems [5].

In a non-Lindblad perspective and on the ground of the path-integral approach to open quantum systems, Caldeira and Leggett [10] derived a Markovian master equation (the so-termed Caldeira—Leggett equation) describing quantum Brownian motion at high temperatures starting from a Hamiltonian modeling the environment as a bath of harmonic oscillators. Such a Caldeira—Leggett equation (CLE) has been extended by Caldeira, Cerdeira, and Ramaswamy [11] for any temperature but for very weak damping. Nevertheless, such Markovian CLEs may give rise to unphysical results, for they are not of the Lindblad form [3—6, 12]. In brief, like the Lindblad master equations, it has been claimed that Markovian CLEs cannot be considered as a *bona fide* description of quantum Brownian motion [8,12], albeit the high-temperature CLE has been employed for accounting for the decoherence process [2,13].

Alternatively, we have derived both Markovian Caldeira—Leggett equations [10,11] *without* alluding to the existence of an underlying Hamiltonian modeling the interaction process between particle and environment. Our non-Hamiltonian approach starts directly from the stochastic dynamics (Langevin and Fokker—Planck equations), thereby giving rise to quantum master equations by means of a quantization method we have called it *dynamical quantization* [5,14].

Thus, in the wake of our dynamical-quantization approach to open quantum systems, the main purpose of the present paper is to explore the physical significance of both non-Markovian and non-equilibrium effects upon the quantum Brownian motion by deriving, on the one hand, a non-Markovian Caldeira—Leggett equation and, on the other hand, a non-Markovian quantum Smoluchowski equation in the absence of inertial force. To this end, we lay out this article as follows: In section 2, we feature the Einstein—Langevin—Kolmogorov approach to open classical systems on the basis of which we can derive both the non-Markovian Klein—Kramers equation and the non-Markovian Smoluchowski equation. In section 3, we arrive at the non-Markovian Caldeira—Leggett equation and study the quantum Brownian motion of a particle in a gravitational field as well as its classical limit. Section 4 addresses the problem of quantizing the non-Markovian Smoluchowski equation and its classical limit, whereas section 5 takes up the quantum tunneling of a non-inertial Brownian particle over a potential barrier. Lastly, summary and outlook are presented in section 6. In addition, three appendices are attached.

## 2. The Einstein—Langevin—Kolmogorov approach

Historically, Einstein [15] accounted for Brownian movement in terms of the time evolution of the probability distribution function (the so-called diffusion equation in configuration space). Then, Langevin [16] alternatively advanced a complementary description of Brownian motion in terms of stochastic differential equations (the so-termed Langevin equations [17]). From the mathematical point of view, Brownian motion has been looked at as a stochastic process within the underpinnings of theory of probability according to Kolmogorov [18]. Such contributions by Einstein, Langevin and Komogorov, we dub them the Einstein—Langevin—Kolmogorov approach to classical open systems.

Section 2.1 considers the Langevin equation as much in the presence as in the absence of inertial force. Sections 2.2 and 2.3 in turn show how the non-Markovian Klein—Kramers and non-Markovian Smoluchowski equation can be derived, respectively.

### *2.1. The Langevin equations*

In order to investigate the erratic motion (the so-called Brownian motion) of a tagged material point immersed in an arbitrary environment (a paradigmatic example of non-Hamiltonian open system), Langevin [16] extended the Newton's deterministic approach by carrying out a random generalization of the Newtonian dynamics through a set of stochastic differential equations (the so-called Langevin equations [5,17] )

$$\frac{dP}{dt} = -\frac{dV(X)}{dX} - 2\gamma P + b\Psi(t), \tag{1}$$

$$\frac{dX}{dt} = \frac{P}{m}, \tag{2}$$

The variables $X = X(t)$ and $P = P(t)$ are kinematical properties inherent to the Brownian particle whereas the variable $\Psi = \Psi(t)$ is intrinsic to the environment. The phenomenological parameter $m$ characterizes the massive structure of the Brownian particle, while $\gamma$ and $b$ are deemed to be positive constants responsible for coupling environment and Brownian particle, so that as $\gamma = b = 0$ the open system (1) and (2) renders isolated from its environment.

In equation (1) the environmental force, given by $F_{\text{env}}(P,\Psi) = -2\gamma P + b\Psi(t)$, is made up by a *memoryless* dissipative force $F_{\text{d}} = -2\gamma P$ responsible for stopping the particle's motion and the so-termed Langevin force $L(t) = b\Psi(t)$ that accounts for activating its movement. Both the phenomenological parameters $b$ and $\gamma$ do control such an environmental influence: $b$ is called fluctuation parameter while $\gamma$ is termed dissipation parameter. The existence of the environmental force $F_{\text{env}}(P,\Psi)$ suggests a relationship between $b$ and $\gamma$. Further, we can readily check that the parameter $b$ may be expressed in dimensions of $[mass \times length \times time^{-3/2}]$, provided that $\Psi(t)$ is in dimensions of $[time^{-1/2}]$.

From the mathematical viewpoint [5,17,18], the quantities $X = X(t)$, $P = P(t)$, and $\Psi = \Psi(t)$ in the Langevin equations (2.3) and (2.4) are viewed as random variables in the sense that there exists a probability distribution function $\mathcal{F}_{XP\Psi}(x,p,\psi,t)$, associated with the whole stochastic system $\{X,P,\Psi\}$, expressed in terms of the possible values $x = \{x_i\}$, $p = \{p_i\}$, and $\psi = \{\psi_i\}$, with $i \geq 1$, distributed about the sharp values $q$, $\overline{p}$ and $\varphi$ of $X$, $P$, and $\Psi$, respectively. In addition, in the Einstein—Langevin—Kolmogorov approach to classical open systems [15,16,18], the average value of any stochastic quantity $A(X,P,\Psi,t)$ is assumed to be expressed as $\langle A(X,P,\Psi,t)\rangle = \iiint_{-\infty}^{+\infty} a(x,p,\psi,t)\mathcal{F}_{XP\Psi}(x,p,\psi,t)dxdpd\psi$, fulfilling the normalization condition $\langle 1\rangle = \iiint_{-\infty}^{+\infty} \mathcal{F}_{XP\Psi}(x,p,\psi,t)dxdpd\psi = 1$.

As the coupling parameters do vanish, i.e., as $b = \gamma = 0$, as well as assuming that $\mathcal{F}_{XP\Psi}(x,p,\psi,t) = \delta(x-q)\delta(p-\overline{p})\delta(\psi-\varphi)$, the Langevin equations (1) and (2) become the deterministic Newton equations in terms of the average values $\langle X\rangle = q$ and $\langle P\rangle = \overline{p}$.

In the absence of inertial force, i.e., as far as the condition $|F_{\text{env}}(P,\Psi)| \gg \left|\frac{dP}{dt}\right|$ is concerned in (1), the focus on the motion of the Brownian particle is gained through the single differential equation

$$\frac{dX}{dt} = -\frac{1}{2m\gamma}\frac{dV(X)}{dX} + \frac{b}{2m\gamma}\Psi(t). \tag{3}$$

That is, the momentum $P$ is said to be eliminated from the description of Brownian movement in the absence of inertial force.

## *2.2. The non-Markovian Klein—Kramers equation*

The Langevin equations (1) and (2) give rise to the Kolmogorov equation in phase space [5,18—20]

$$\frac{\partial \mathcal{F}(x,p,t)}{\partial t} = \mathbb{K}\mathcal{F}(x,p,t), \tag{4}$$

where the Kolmogorovian operator $\mathbb{K}$ acts upon the (marginal) probability distribution function $\mathcal{F}(x,p,t) = \iint_{-\infty}^{\infty} \mathcal{F}_{XP\Psi}(x,p,\psi,t)d\psi$ according to

$$\mathbb{K}\mathcal{F}(x,p,t) = \sum_{k=1}^{\infty}\sum_{r=0}^{k}\frac{(-1)^k}{r!\,(k-r)!}\frac{\partial^k}{\partial x^{k-r}\partial p^r}\left[A^{(k-r,r)}(x,p,t)\mathcal{F}(x,p,t)\right] \qquad (5)$$

with

$$A^{(k-r,r)}(x,p,t) = \lim_{\varepsilon\to 0}\left[\frac{\langle(\Delta X)^{k-r}\rangle\langle(\Delta P)^r\rangle}{\varepsilon}\right]. \qquad (6)$$

In (6) both increments $\Delta X = X(t+\varepsilon) - X(t)$ and $\Delta P = P(t+\varepsilon) - P(t)$ are to be calculated from (1) and (2) from the integral equations $\Delta X = \frac{1}{m}\int_t^{t+\varepsilon} P(t)dt$ and $\Delta P = -\frac{dV(X)}{dX}\varepsilon - 2\gamma\int_t^{t+\varepsilon} P(t)dt + b\int_t^{t+\varepsilon}\Psi(t)dt$. Moreover, the average values, $\langle(\Delta X)^{k-r}\rangle\langle(\Delta P)^r\rangle$, are to be calculated about the sharp values $q$ and $\overline{p}$, i.e., $\mathcal{F}_{XP\Psi}(x,p,\psi,t) = \delta(x-q)\delta(p-\overline{p})\mathcal{F}_\Psi(\psi,t)$.

The phase-space Kolmogorov equation (4) describes the time evolution of a Brownian particle immersed in a non-Gaussian environment since the coefficients (6) rely on the infinite moments of the Langevin force: $\langle L(t_1)\dots L(t_k)\rangle$, with $k = 1,2,\dots,\infty$. According to the Pawula theorem [20], there exists no non-Gaussian approximation to (4) complying with the positivity of $\mathcal{F}(x,p,t)$. A sufficient condition leading to a Gaussian Kolmogorov equation (4) is then to consider $|x_2 - x_1|^3 \ll 0$, where $x_2 \equiv x(t+\varepsilon)$ and $x_1 \equiv x(t)$ are associated with the random variables $X(t+\varepsilon)$ and $X(t)$, respectively.

So, taking into account $\lim_{\varepsilon\to 0}\int_t^{t+\varepsilon}\langle\Psi(t')\rangle dt' = 0$, the Brownian motion of a particle immersed in a generic stationary environment turns out to be described by the Langevin equations (1) and (2) and the corresponding Gaussian Kolmogorov equation (the so-called Fokker—Planck equation)

$$\frac{\partial\mathcal{F}}{\partial t} = -\frac{p}{m}\frac{\partial\mathcal{F}}{\partial x} + \frac{\partial}{\partial p}\left[\frac{\partial\mathcal{V}_{\text{eff}}(x,t)}{\partial x} + 2\gamma p\right]\mathcal{F} + 2\gamma m\mathcal{E}(t)\frac{\partial^2\mathcal{F}}{\partial p^2}, \qquad (7)$$

where $\mathcal{F} \equiv \mathcal{F}(x,p,t)$ and $\mathcal{V}_{\text{eff}}(x,t)$ the effective potential

$$\mathcal{V}_{\text{eff}}(x,t) = V(x) - x\sqrt{4\gamma m\mathcal{E}(\infty)}\langle\Psi(t)\rangle \qquad (8)$$

with the mean value of $\Psi(t)$ given by $\langle\Psi(t)\rangle = \lim_{\varepsilon\to 0}\frac{1}{\varepsilon}\int_t^{t+\varepsilon}\langle\Psi(t')\rangle dt'$. Further, the time-dependent diffusion coefficient $\mathcal{D}(t) = 2\gamma m\mathcal{E}(t)$ is expressed in terms of the function

$$\mathcal{E}(t) = \frac{b^2}{4\gamma m}I(t), \qquad (9)$$

where $I(t)$ is the dimensionless function

$$I(t) = \lim_{\varepsilon \to 0} \frac{1}{\varepsilon} \iint_{t}^{t+\varepsilon} \langle \Psi(t_1) \Psi(t_2) \rangle dt_1 dt_2. \qquad (10)$$

It is readily to check that the time-dependent function $\mathcal{E}(t)$ defined by (9) has dimensions of energy, i.e., $[mass \times length^2 \times time^{-2}]$. Hence we call it the diffusion energy responsible for the Brownian motion of the particle immersed in a generic Gaussian environment.

The Fokker—Planck equation (7) is to be solved from a given initial condition $\mathcal{F}(x, p, t = 0)$ evolving towards a steady solution $\lim_{t \to \infty} \mathcal{F}(x, p, t) \approx \mathcal{F}(x, p)$. In this steady regime the diffusion energy (9) displays the asymptotic behavior $\lim_{t \to \infty} \mathcal{E}(t) \approx \mathcal{E}(\infty) = \frac{b^2}{4\gamma m}$, provided that $\lim_{t \to \infty} I(t) \approx I(\infty) = 1$. The physical significance of this long-time behavior has to do with the fact that environmental fluctuations do possess Markovian correlations, i.e., the correlational function $I(t)$ exhibits a local (short) feature decaying to one in the stationary regime. By contrast, non-Markovian effects show up in the non-equilibrium time window $0 < t < \infty$. Moreover, the steady diffusion energy $\mathcal{E}(\infty) = \frac{b^2}{4\gamma m}$ yields the Markovian fluctuation-dissipation relation in the form $b = \sqrt{4\gamma m \mathcal{E}(\infty)}$, meaning that the fluctuation coefficient $b$ can be determined in terms of the friction constant $\gamma$, the mass of the particle $m$, as well as $\mathcal{E}(\infty)$.

The characteristic feature underlying the concept of time-dependent diffusion energy (9), $\mathcal{E}(t) = \mathcal{E}(\infty) I(t)$, is that it sets up a general relationship between fluctuation and dissipation processes as well as fulfilling the validity condition $0 < \mathcal{E}(t) < \infty$. Both cases $\mathcal{E}(t) = 0$ and $\mathcal{E}(t) = \infty$ should be disregarded, for they may violate the fluctuation—dissipation relation: the former case may lead to dissipation without fluctuation, while the latter one may give rise to fluctuation without dissipation.

If $\langle \Psi(t) \rangle = 0$, $I(t) = 1$, and the environment is deemed to be a heat bath in thermodynamic equilibrium at temperature $T$, so that we can identify the steady diffusion energy $\mathcal{E}(\infty)$ with the thermal energy $k_B T$, where $k_B$ is the Boltzmann constant (see Appendix A), then the non-Markovian Fokker—Planck equation (7) reduces to the Markovian Klein—Kramers equation [5,17,21]. Hence, we term (7) the non-Markovian Klein—Kramers equation.

## *2.3. The non-Markovian Smoluchowski equation*

The Langevin equation (3) gives rise to the following Kolmogorov equation in configuration space [5,18—20]

$$\frac{\partial f(x,t)}{\partial t} = \sum_{k=1}^{\infty} \frac{(-1)^k}{k!} \frac{\partial^k}{\partial x^k} [A_k(x,t) f(x,t)], \qquad (11)$$

with $f(x,t) = \int_{-\infty}^{\infty} \mathcal{F}_{X\Psi}(x,\psi,t) d\psi$. The coefficients $A_k(x,t)$ being given by $A_k(x,t) = \lim_{\varepsilon \to 0} \left[\frac{\langle (\Delta X)^k \rangle}{\varepsilon}\right]$, where the average values, $\langle (\Delta X)^k \rangle$, are to be calculated about the sharp values $q$, i .e., $\mathcal{F}_{X\Psi}(x,\psi,t) = \delta(x-q)\mathcal{F}_{\Psi}(\psi,t)$ from the Langevin equation (3) in the integral form: $\Delta X \equiv X(t+\varepsilon) - X(t) = -\frac{\varepsilon}{2m\gamma}\frac{dV}{dX} + \sqrt{\frac{\mathcal{E}(\infty)}{m\gamma}} \int_t^{t+\varepsilon} \Psi(t')dt'$. Accordingly, the configuration-space Kolmogorov equation (11) in the Gaussian approximation, $|x_2 - x_1|^3 \ll 0$, reduces to the following non-Markovian Fokker—Planck equation

$$\frac{\partial f(x,t)}{\partial t} = \frac{1}{2m\gamma} \frac{\partial}{\partial x} \left[ \frac{\partial \mathcal{V}_{\text{eff}}(x,t)}{\partial x} f(x,t) \right] + \frac{\mathcal{E}(\infty)}{2m\gamma} I(t) \frac{\partial^2 f(x,t)}{\partial x^2}, \qquad (12)$$

where $D(t) = \frac{\mathcal{E}(\infty)}{2m\gamma} I(t)$ is the time-dependent diffusion coefficient and $\mathcal{V}_{\text{eff}}(x,t)$ the effective potential (8).

For the case in which $\langle \Psi(t) \rangle = 0$, $I(t) = 1$ as well as $\mathcal{E}(\infty) = k_B T$ the Fokker—Planck equation (12) becomes the Markovian Smoluchowski equation [5,17,22]. For this reason, we dub (12) the non-Markovian Smoluchowski equation.

We wish to point out that both the Markovian Klein—Kramers equation [21] and the Markovian Smoluchowski equation [22] can be obtained as a special case from (7) and (12), respectively, without postulating *ab initio* both the statistical properties $\langle L(t) \rangle = 0$ $\langle L(t)L(t') \rangle = 4\gamma m k_B T \delta(t-t')$ of the Langevin force $L(t)$ as is commonly achieved in the literature [16,17,22]. By contrast, in our approach the important fact is the asymptotic behavior of $\langle \Psi(t) \rangle$ as well as $I(t)$ in the stationary limit. In other words, both the statistical properties above turn up as sufficient but not necessary conditions for attaining the thermodynamic equilibrium state.

Having showed how non-Markovian effects can turn up in both Klein—Kramers and Smoluchowski equations, in the next sections 3 and 4 we intend to examine how quantum effects upon these kinds of non-Markovian Fokker—Planck equations can arise via a non-Hamiltonian quantization process.

## 3. Deriving the non-Markovian Caldeira—Leggett equation

Having derived the Kolmogorov equation in phase space (4) for a Brownian particle immersed in a non-Gaussian environment, we now wish to quantize it by means of a non-Hamiltonian quantization process called dynamical quantization [5,14]. First, we obtain the equation of motion

$$\frac{\partial \chi(x,\eta,t)}{\partial t} = \int_{-\infty}^{\infty} \mathbb{K}\mathcal{F}(x,p,t)e^{ip\eta}dp, \tag{13}$$

after performing upon (4) the Fourier transform $\chi(x,\eta,t) = \int_{-\infty}^{\infty} \mathcal{F}(x,p,t)e^{ip\eta}dp$, where the exponential $e^{ip\eta}$ is deemed to be a dimensionless term. Then, the Kolmogorov stochastic dynamics (4) is said to be quantized by introducing into equation (13) the following quantization conditions

$$x_1 = x + \frac{\eta\hbar}{2} \qquad ; \qquad x_2 = x - \frac{\eta\hbar}{2}, \tag{14}$$

whereby the transformation parameter ħ from the change of variables $(x,\eta) \mapsto (x_1,x_2)$ is dubbed Planck's constant having dimensions of angular momentum, i.e., $[mass \times length^2 \times time^{-1}]$. The geometric meaning of the quantization conditions (14) lies at the existence of a minimal distance between the points $x_1$ and $x_2$ on account of the quantum nature of space, i.e., $|x_2 - x_1| = |\eta\hbar|$, such that in the classical limit $\hbar \to 0$, physically interpreted as $|\eta\hbar| \ll |x_2 - x_1|$, the result $x_2 = x_1 = x$ can be readily recovered.

Thus, after inserting both quantization conditions (14) into (13) we arrive at the non-Gaussian quantum master equation

$$\frac{\partial \rho(x_1,x_2,t)}{\partial t} = \int_{-\infty}^{\infty} \mathbb{K}\mathcal{F}\left(\frac{x_1 - x_2}{2}, p, t\right) e^{ip\frac{(x_1-x_2)}{\hbar}}dp, \tag{15}$$

which describes the quantum Brownian motion of a particle in the presence of a generic quantum fluid, for instance.

On going from the classical equation of motion (13) to the quantum dynamics (15) via quantization conditions (14), we have replaced the classical function $\chi \equiv \chi(x,\eta,t)$ with the quantum function $\rho \equiv \rho(x_1,x_2,t)$, since it turns out to depend on the Planck constant, ħ. We dub $\rho(x_1,x_2,t)$ the von Neumann function.

Taking the Gaussianity condition $|x_2 - x_1|^3 \ll 0$ into account, our non-Gaussian quantum master equation (15) reduces to

$$i\hbar\frac{\partial\rho}{\partial t}=\left[\mathcal{V}_{\text{eff}}^{(\hbar)}(x_1,t)-\mathcal{V}_{\text{eff}}^{(\hbar)}(x_2,t)\right]\rho-\frac{\hbar^2}{2m}\left(\frac{\partial^2\rho}{\partial x_1^2}-\frac{\partial^2\rho}{\partial x_2^2}\right)-i\hbar\gamma(x_1-x_2)\left(\frac{\partial\rho}{\partial x_1}-\frac{\partial\rho}{\partial x_2}\right)$$
$$-\frac{i\mathcal{D}_\hbar(t)}{\hbar}(x_1-x_2)^2\rho, \tag{16}$$

where the effective potentials $\mathcal{V}_{\text{eff}}^{(\hbar)}(x_k,t)$, with $k=1{,}2$, are given by $\mathcal{V}_{\text{eff}}^{(\hbar)}(x_k,t)=V(x_k)-x_k\sqrt{4\gamma m\mathcal{E}_\hbar(\infty)}\langle\Psi(t)\rangle$ and $\mathcal{D}_\hbar(t)$ is the time-dependent quantum diffusion coefficient $\mathcal{D}_\hbar(t)=2\gamma m\mathcal{E}_\hbar(\infty)I(t)$.

It is relevant to notice that in the Gaussian quantum master equation (16), the time evolution parameter $t$, the mass $m$, the frictional constant $\gamma$, as well as the both functions $\langle\Psi(t)\rangle$ and $I(t)$ are deemed to be non-quantized quantities, that is, they are ħ-independent, whereas the classical diffusion energy $\mathcal{E}(\infty)$ has been subject to a quantization process, i.e., $\mathcal{E}(\infty)\to\mathcal{E}_\hbar(\infty)$.

The master equation (16) describes the quantum Brownian motion of a particle immersed in a generic Gaussian quantum environment. If, for the specific case of a heat bath consisting of quantum harmonic oscillators with oscillation frequency $\omega$, we could identify the Brownian particle's steady diffusion energy $\mathcal{E}_\hbar(\infty)$ with the mean thermal energy of the bath in thermodynamic equilibrium, then we obtain $\mathcal{E}_\hbar(\infty)=\frac{b_\hbar^2}{4\gamma m}=\frac{\omega\hbar}{2}\coth\left(\frac{\omega\hbar}{2k_BT}\right)$ (see Appendix A). Accordingly, the equilibrium quantum fluctuation—dissipation relation reads $b_\hbar=\sqrt{2\gamma m\omega\hbar\coth\left(\frac{\omega\hbar}{2k_BT}\right)}$ reducing to $b=\sqrt{4\gamma m\mathcal{E}(\infty)}$ in the classical limit, i.e., as $k_BT\gg\omega\hbar/2$. Moreover, the quantum diffusion constant $\mathcal{D}_\hbar(t)$ in (16) reads $\mathcal{D}_\hbar(t)=\gamma m\omega\hbar\coth\left(\frac{\omega\hbar}{2k_BT}\right)I(t)$, yielding at high temperatures, $T\gg\omega\hbar/2k_B$, the classical diffusion constant in (7) for thermal systems: $\mathcal{D}(t)=2\gamma mk_BTI(t)$. On the other hand, the zero-point diffusion constant reads $\mathcal{D}_\hbar^{(T=0)}(\infty)=\gamma m\omega\hbar$. Thus, for thermal quantum systems, the non-Markovian quantum master equation (16) reads

$$i\hbar\frac{\partial\rho}{\partial t}=\left[\mathcal{V}_{\text{eff}}^{(\hbar)}(x_1,t)-\mathcal{V}_{\text{eff}}^{(\hbar)}(x_2,t)\right]\rho-\frac{\hbar^2}{2m}\left(\frac{\partial^2\rho}{\partial x_1^2}-\frac{\partial^2\rho}{\partial x_2^2}\right)$$
$$-i\hbar\gamma(x_1-x_2)\left(\frac{\partial\rho}{\partial x_1}-\frac{\partial\rho}{\partial x_2}\right)-i\gamma m\omega\coth\left(\frac{\omega\hbar}{2k_BT}\right)I(t)(x_1-x_2)^2\rho \tag{17}$$

which describes a Brownian particle subject to averaging effects present in the effective potentials $\mathcal{V}_{\text{eff}}^{(\hbar)}(x_k,t)$ as well as non-Markovian effects through the correlational function $I(t)$. In addition, it is worth underscoring that (17) has been derived for any initial condition $\rho(x_1,x_2,t=0)$.

As long as $\langle\Psi(t)\rangle=0$ and $I(t)=1$, our quantum master equation (17) leads to the Markovian Caldeira—Leggett equation

$$i\hbar\frac{\partial\rho}{\partial t} = [V(x_1,t) - V(x_2,t)]\,\rho - \frac{\hbar^2}{2m}\left(\frac{\partial^2\rho}{\partial x_1^2} - \frac{\partial^2\rho}{\partial x_2^2}\right)$$

$$-i\hbar\gamma(x_1 - x_2)\left(\frac{\partial\rho}{\partial x_1} - \frac{\partial\rho}{\partial x_2}\right) - i\gamma m\omega \coth\left(\frac{\omega\hbar}{2k_B T}\right)(x_1 - x_2)^2\rho \qquad (18)$$

found out by Caldeira, Cerdeira, and Ramaswamy [11] following the Feynman path integral formalism and making assumptions on the weakness of the damping $\gamma \ll \omega$. Yet our derivation has shown that Feynman's formalism as well as such an assumption are thoroughly unnecessary features for reaching (18). Moreover, on the condition that the thermal reservoir holds at high temperatures, i.e., $\coth(\omega\hbar/2k_B T) \sim 2k_B T/\omega\hbar$, our master equation (17), along with $\langle\Psi(t)\rangle = 0$ and $I(t) = 1$, yields the following master equation

$$i\hbar\frac{\partial\rho}{\partial t} = [V(x_1,t) - V(x_2,t)]\,\rho - \frac{\hbar^2}{2m}\left(\frac{\partial^2\rho}{\partial x_1^2} - \frac{\partial^2\rho}{\partial x_2^2}\right)$$

$$-i\hbar\gamma(x_1 - x_2)\left(\frac{\partial\rho}{\partial x_1} - \frac{\partial\rho}{\partial x_2}\right) - \frac{2i\gamma m k_B T}{\hbar}(x_1 - x_2)^2\rho \qquad (19)$$

originally found by Caldeira and Leggett [10] upon making use of the path integral techniques and assuming that particle and environment are initially uncorrelated. This assumption is non-realistic and leads to non-physical results [12,23]. Our approach on the contrary has shown that the high-temperature Markovian Caldeira—Leggett equation (19) can be derived for any initial condition as long as averaging and non-Markovian effects could be neglected in our master equation (17) at high temperatures.

In addition, it has been claimed that the Markovian Caldeira—Leggett equation (19) may also lead to unphysical results because it is not of the Lindblad form [12]. To overcome this difficulty, terms has been added to (19) for fitting with the Lindblad framework. Yet, according to our approach such an *ad hoc* procedure is unnecessary given that our quantum master equation (17) conveys non-Markovian effects, so bypassing Lindblad's condition.

Lastly, because our non-Markovian quantum master equation (17) contains both the Markovian Caldeira—Leggett equations (18) and (19) as special cases, we dub it the non-Markovian Caldeira—Leggett equation.

### *3.1. A quantum Brownian particle in a gravitational field*

In order to provide physical significance to our non-Markovian Caldeira—Leggett equation (17), let us consider the correlational function of the form $I(t) = 1 - e^{-t/t_c}$ (see Appendix B) and the case of the quantum Brownian motion of a particle of mass $m$ moving due to the gravitational energy, $V(x) = mgx$, where $g$ denotes the value of gravity acceleration. This free-falling quantum Brownian motion is described by the following dynamics in quantum phase space

$$\frac{\partial W}{\partial t} = -\frac{1}{m}\left[p - \frac{mg}{2\gamma} + \sqrt{\frac{m\mathcal{E}_{\hbar}(\infty)}{\gamma}}\langle\Psi(t)\rangle\right]\frac{\partial W}{\partial x} + 2\gamma\frac{\partial}{\partial p}[pW] + \mathcal{D}_{\hbar}(t)\frac{\partial^2 W}{\partial p^2}, \qquad (20)$$

in terms of the effective momentum $p = p' + \frac{mg}{2\gamma} - \sqrt{\frac{m\mathcal{E}_{\hbar}(\infty)}{\gamma}}\langle\Psi(t)\rangle$ and the time-dependent quantum diffusion coefficient $\mathcal{D}_{\hbar}(t) = 2\gamma m\mathcal{E}_{\hbar}(\infty)\left(1 - e^{-\frac{t}{t_c}}\right)$. The function $W \equiv W(x, p, t)$ in (20) is the Wigner transform

$$W(x,p,t) = \frac{1}{2\pi}\int_{-\infty}^{\infty}\rho(x + \frac{\eta\hbar}{2}, x - \frac{\eta\hbar}{2}, t)e^{-ip\eta}d\eta \qquad (21)$$

performed upon the von Neuman function in the master equation (17).

To solve equation (20) we start with the following initial condition that couples Brownian particle with the environment

$$W(x,p,t=0) = \frac{1}{\pi\hbar}e^{-\left(\frac{\tau_r p^2}{m\hbar} + \frac{mx^2}{\hbar\tau_r}\right)}, \qquad (22)$$

where $\tau_r = (4\gamma)^{-1}$ denotes a sort of relaxation time. It is straightforward to check that (22) complies with the Heisenberg fluctuation relation $\Delta P(0)\Delta X(0) = \hbar/2$.

Taking $W(p,t) = \int_{-\infty}^{\infty} W(x,p,t)dx$ into account, equation (20) turns out to be rewritten as the quantum Rayleigh equation

$$\frac{\partial W(p,t)}{\partial t} = 2\gamma\frac{\partial}{\partial p}[pW(p,t)] + \mathcal{D}_{\hbar}(t)\frac{\partial^2 W(p,t)}{\partial p^2} \qquad (23)$$

the time-dependent solution of which reads

$$W(p,t) = \frac{1}{\sqrt{4\pi A_{\hbar}(t)}}e^{-\frac{p^2}{4A_{\hbar}(t)}} \qquad (24)$$

with

$$A_{\hbar}(t) = \frac{m\hbar}{4\tau_r}e^{-\frac{t}{\tau_r}} + \frac{m\omega\hbar}{4}\coth\left(\frac{\omega\hbar}{2k_B T}\right)\left[1 - e^{-\frac{t}{\tau_r}} + \frac{t_c}{(t_c - \tau_r)}\left(e^{-\frac{t}{\tau_r}} - e^{-\frac{t}{t_c}}\right)\right]. \qquad (25)$$

The probability distribution function (24) leads to $\langle P\rangle = 0$ and $\langle P^2\rangle = 2A_{\hbar}(t)$. Hence, the momentum fluctuation $\Delta P(t) = \sqrt{\langle P^2\rangle - \langle P\rangle^2}$ reads

$$\Delta P(t) = \sqrt{\frac{m\hbar}{2\tau_r}e^{-\frac{t}{\tau_r}} + \frac{m\omega\hbar}{2}\coth\left(\frac{\omega\hbar}{2k_B T}\right)\left[1 - e^{-\frac{t}{\tau_r}} + \frac{t_c}{(t_c - \tau_r)}\left(e^{-\frac{t}{\tau_r}} - e^{-\frac{t}{t_c}}\right)\right]}. \qquad (26)$$

Because $\Delta P(t) = \Delta P'(t)$ we conclude that the mean value of $\Psi$ as well as the gravity effect present in the quantum Klein—Kramers equation (20) are unobservable. Further, it is readily to check that non-Markovian effects account for diminishing the strength of the fluctuation (26).

From (24) in the steady regime we can derive

$$W(p) = \frac{1}{\sqrt{\pi m\omega\hbar\coth\left(\frac{\omega\hbar}{2k_BT}\right)}} e^{-\frac{p^2}{m\omega\hbar\coth\left(\frac{\omega\hbar}{2k_BT}\right)}}, \tag{27}$$

meaning that both relaxation and non-Markovian effects in (24) die out in thermal equilibrium. In other words, both relaxation and non-Markovian features in (24) are intrinsically non-equilibrium effects.

Furthermore, the root mean square momentum (26) gives rise to the following thermal quantum force

$$\mathcal{K}(t) = \frac{d\Delta P(t)}{dt} = -2\gamma\sqrt{2\gamma\hbar m}\mathrm{B}(t) \tag{28}$$

with the dimensionless function $\mathrm{B}(t)$

$$\mathrm{B}(t) = \frac{\tau_r^2\omega\coth\left(\frac{\omega\hbar}{2k_BT}\right)\left(e^{-\frac{t}{\tau_r}} - e^{-\frac{t}{t_c}}\right) + (t_c - \tau_r)e^{-\frac{t}{\tau_r}}}{\sqrt{(t_c - \tau_r)\left\{(t_c - \tau_r)e^{-\frac{t}{\tau_r}} + \tau_r\omega\coth\left(\frac{\omega\hbar}{2k_BT}\right)\left[\tau_r\left(e^{-\frac{t}{\tau_r}} - 1\right) + t_c\left(1 - e^{-\frac{t}{t_c}}\right)\right]\right\}}} \tag{29}$$

displaying the following physically accessible time scales: the evolution time $t$ (the observation time, for instance), the correlation time $t_c$, the relaxation time $\tau_r$, the oscillation time $t_{\mathrm{osc}} = \omega^{-1}$, as well as the quantum time $t_{\mathrm{q}} = \hbar/k_BT$. If $\mathrm{B}(t) > 0$, then the force (28) is said to be attractive: $\mathcal{K}(t) < 0$. Otherwise, if $\mathrm{B}(t) < 0$, then (28) renders repulsive, i.e., $\mathcal{K}(t) > 0$.

The quantum force (28) is a non-equilibrium effect since it vanishes in the steady regime, whereas at short times $t \ll t_c, t_r$ the force (28) becomes attractive with the constant value $\mathcal{K}(t) = -2\gamma\sqrt{2\gamma\hbar m}$. This fact implies that the root mean square momentum (26) is a differentiable quantity for all time $t \geq 0$.

By way of example, we take $|\mathrm{B}(t)|\sim 1$, reckon with the Munro and Gardiner's values for the parameters $\gamma$, $\hbar$, and $m$ in Ref. [12], i.e., $\gamma\sim 10^{11}\mathrm{s}^{-1}$, $\hbar\sim 10^{-34}\mathrm{m}^2\mathrm{kgs}^{-1}$, and $m\sim 10^{-26}\mathrm{kg}$, and then obtain the magnitude $|\mathcal{K}(t)|\sim 10^{-13}\mathrm{N}$.

This simple numerical example suggests that the strength of the quantum thermal force (28) could be measured in experiences, for instance, using trapped ions [24] in which measurement of forces of order of yoctonewton, i.e., $10^{-24}$N, has been recently reported.

At zero temperature, i.e., $\coth(\omega\hbar/2k_BT) \sim 1$, the dimensionless factor (29) becomes

$$\mathrm{B}^{(T=0)}(t) = \frac{\tau_r^2\omega\left(e^{-\frac{t}{t_r}} - e^{-\frac{t}{t_c}}\right) + (t_c - \tau_r)e^{-\frac{t}{\tau_r}}}{\sqrt{(t_c - \tau_r)\left\{(t_c - \tau_r)e^{-\frac{t}{\tau_r}} + \omega\tau_r\left[\tau_r\left(e^{-\frac{t}{\tau_r}} - 1\right) + t_c\left(1 - e^{-\frac{t}{t_c}}\right)\right]\right\}}}, \quad (30)$$

while at high temperatures, i.e., as $\coth(\omega\hbar/2k_BT) \sim 2k_BT/\omega\hbar$, it turns out to be

$$\mathrm{B}(t) = \frac{2k_BT\tau_r^2\left(e^{-\frac{t}{\tau_r}} - e^{-\frac{t}{t_c}}\right) + \hbar(t_c - \tau_r)e^{-\frac{t}{\tau_r}}}{\sqrt{\hbar(t_c - \tau_r)\left\{\hbar(t_c - \tau_r)e^{-\frac{t}{\tau_r}} + 2k_BT\tau_r\left[\tau_r\left(e^{-\frac{t}{\tau_r}} - 1\right) + t_c\left(1 - e^{-\frac{t}{t_c}}\right)\right]\right\}}}, \quad (31)$$

thereby suggesting that the non-equilibrium quantum effect (28) could show up at both high and zero temperatures.

### *3.2. Classical limit*

The classical limit $\hbar \to 0$ should be physically interpreted as the classical thermal energy is deemed to be too large in comparison to the quantal energy, i.e., $k_BT \gg \frac{\omega\hbar}{2}$, so that $\mathcal{E}_\hbar(\infty) = \frac{\omega\hbar}{2}\coth\left(\frac{\omega\hbar}{2k_BT}\right) \sim \mathcal{E}(\infty) = k_BT$. So, from the quantum Wigner function (24) in the steady regime at high temperatures, we can derive the well-known Maxwell—Boltzmann probability distribution function $f(p) = \frac{1}{\sqrt{2\pi m k_BT}}e^{-\frac{p^2}{2mk_BT}}$ for the effective momentum $p = p' + \frac{mg}{2\gamma} - \sqrt{\frac{mk_BT}{\gamma}}\langle\Psi(t)\rangle$, while in the classical limit the sort of quantum force (28) becomes

$$F(t) = -2\gamma\sqrt{mk_BT}\,\mathrm{G}(t) \quad (32)$$

with the dimensionless function $\mathrm{G}(t)$ being

$$\mathrm{G}(t) = \frac{t_r\left(e^{-\frac{t}{t_r}} - e^{-\frac{t}{t_c}}\right)}{\sqrt{(t_c - t_r)\left[t_r\left(e^{-\frac{t}{t_r}} - 1\right) + t_c\left(1 - e^{-\frac{t}{t_c}}\right)\right]}}. \quad (33)$$

Taking into account $|G(t)| \sim 1$, $\gamma \sim 10^{11} s^{-1}$, $m \sim 10^{-26}$kg, $k_B \sim 10^{-23} m^2 kgs^{-2} K^{-1}$, and $T \sim 1000$K, we find $|F(t)| \sim 10^{-22}$N, which is the magnitude of the classical thermal force (32) exerted by a heat bath at 1000K upon the Brownian motion of a free-falling particle of mass $10^{-26}$kg .

The dimensionless function (33) approximates to $G(t) \sim -(2\gamma t_c)^{-1/2}$ at short times. Hence the non-equilibrium classical thermal force (32) reads $F(t) = \sqrt{\frac{2\gamma m k_B T}{t_c}}$ which blows up in the Markovian limit $t_c \to 0$. This upshot reveals the pivotal importance of non-Markovian effects for the existence of the physical concept of non-equilibrium thermal force (32) in the classical realm. In mathematical parlance, non-Markovian features account for the differentiability property of the root mean square momentum $\Delta P(t) = \sqrt{\langle P^2 \rangle - \langle P \rangle^2}$ for all time $t$ in the classical domain. Accordingly, the Langevin equation

$$\frac{dP}{dt} = -2\gamma P + \sqrt{4\gamma m k_B T}\Psi(t),$$

underlying the non-Markovian classical Rayleigh equation

$$\frac{\partial f(p,t)}{\partial t} = 2\gamma \frac{\partial}{\partial p}[p f(p,t)] + 2\gamma m k_B T I(t) \frac{\partial^2 f(p,t)}{\partial p^2},$$

obtained as classical limit of (23), should be interpreted as a differential stochastic equation and not as an integral equation according to a determined rule of interpretation (Doob's interpretation, for instance. See Coffey et al.'s book in Ref. [17] as well as [25]).

### *3.3. Discussions*

The predominant paradigm for describing quantum Brownian motion can be summarized as follows [5,6,8]: Envisage an environment consisting of a set of harmonic oscillators coupled to a Brownian particle; having built up the Hamiltonian or Lagrangian function of the whole system (particle plus environment), canonical quantization procedure (Dirac or Heisenberg quantization) could be employed to enter into quantum world. Once established this quantum Hamiltonian picture portraying dissipation along with fluctuation, Feynman's path integral formalism or projection operator techniques may be then invocated for deriving master equations for the density matrix describing the quantum motion of the Brownian particle, after getting rid of the medium's variables. On the basis of the former procedure, Caldeira and Leggett [10] quantized the Markovian Klein—Kramers equation [21] and obtained the so-called Markovian Caldeira—Leggett equation. Yet, this quantum master equation is plagued with the problem of positivity for it is not of the Lindblad form [2].

By contrast, according to our dynamical-quantization approach to quantum Brownian motion we have derived the non-Gaussian Kolmogorov quantum master equation (15) without building up a Hamiltonian model for the isolated system (Brownian particle plus environment). Our quantization method is intrinsically non-Hamiltonian in full. By the same token, we reckon that our Caldeira—Leggett quantum master equation (17) eschews the issue of complying with the Lindblad requirement because it gets the non-Markovian gist of the interaction between the Brownian particle and the thermal reservoir in the Gaussian approximation.

## 4. The non-Markovian quantum Smoluchowski equation

In order to provide some generality to our method of non-Hamiltonian quantization, we turn to quantize the non-Markovian Smoluchowski equation (12) describing a harmonic oscillator. The most general case given by the quantization of the Kolmogorov equation in configuration space (11) is taken up in Appendix C.

### *4.1. The harmonic oscillator*

We start with (12) for $V(x) = kx^2/2$ at points $x_1$ and $x_2$, i.e.,

$$\frac{\partial \chi(x_1,t)}{\partial t} = \frac{1}{2m\gamma}\frac{\partial}{\partial x_1}\left[\frac{\partial \mathcal{V}_{\text{eff}}(x_1,t)}{\partial x_1}\chi(x_1,t)\right] + \overline{\mathcal{D}}(t)\frac{\partial^2 \chi(x_1,t)}{\partial {x_1}^2}, \qquad (34)$$

$$\frac{\partial \chi(x_2,t)}{\partial t} = \frac{1}{2m\gamma}\frac{\partial}{\partial x_2}\left[\frac{\partial \mathcal{V}_{\text{eff}}(x_2,t)}{\partial x_2}\chi(x_2,t)\right] + \overline{\mathcal{D}}(t)\frac{\partial^2 \chi(x_2,t)}{\partial {x_2}^2}, \qquad (35)$$

where $\mathcal{V}_{\text{eff}}(x_k,t)$, with $k = 1,2$, are the effective potentials $\mathcal{V}_{\text{eff}}(x_k,t) = \frac{kx_k^2}{2} - 2x_k\sqrt{2m\gamma\mathcal{E}(\infty)}\langle\Psi(t)\rangle$, whereas $\overline{\mathcal{D}}(t)$ is the time-dependent diffusion coefficient $\overline{\mathcal{D}}(t) = 2\mathcal{D}(t) = \frac{\mathcal{E}(\infty)}{m\gamma}I(t)$ associated with both solutions $\chi(x_1,t) = \sqrt{f(x_1,t)}$ and $\chi(x_2,t) = \sqrt{f(x_2,t)}$.

Multiplying (34) and (35) by $\chi(x_2,t)$ and $\chi(x_1,t)$, respectively, and then adding the resulting equations we arrive at

$$\frac{\partial \xi}{\partial t} = \frac{1}{2m\gamma}\left\{\frac{\partial}{\partial x_1}\left[\frac{\partial \mathcal{V}_{\text{eff}}(x_1,t)}{\partial x_1}\xi\right] + \frac{\partial}{\partial x_2}\left[\frac{\partial \mathcal{V}_{\text{eff}}(x_2,t)}{\partial x_2}\xi\right]\right\} + \overline{\mathcal{D}}(t)\left[\frac{\partial^2 \xi}{\partial {x_1}^2} + \frac{\partial^2 \xi}{\partial {x_2}^2}\right], \qquad (36)$$

where $\xi \equiv \xi(x_1,x_2,t) = \chi(x_1,t)\,\chi(x_2,t)$. By quantizing via (14) in the form $x_1 = x' + \frac{\eta\hbar}{2}$ and $x_2 = x' - \frac{\eta\hbar}{2}$, as well as making use of the following relations $\frac{\partial}{\partial x_1} = \frac{1}{2}\frac{\partial}{\partial x'} - \frac{1}{\hbar}\frac{\partial}{\partial \eta}$ and $\frac{\partial}{\partial x_2} = \frac{1}{2}\frac{\partial}{\partial x'} + \frac{1}{\hbar}\frac{\partial}{\partial \eta}$, the classical equation of motion (36) changes into the quantum equation of motion

$$\frac{\partial \rho}{\partial t} = \frac{k}{m\gamma}\rho + \frac{kx}{2m\gamma}\frac{\partial \rho}{\partial x} + \frac{k}{2m\gamma}\eta\frac{\partial \rho}{\partial \eta} + \frac{\mathcal{E}_\hbar(t)}{2m\gamma}\left[\frac{\partial^2 \rho}{\partial x^2} + \frac{4}{\hbar^2}\frac{\partial^2 \rho}{\partial \eta^2}\right] \qquad (37)$$

expressed in terms of the effective position $x = x' - \frac{2}{k}\sqrt{2m\gamma\mathcal{E}_\hbar(\infty)}\langle\Psi(t)\rangle$.

Upon going from (36) to (37) we have replaced the solution $\xi = \xi(x_1, x_2, t)$ with the ħ-dependent function $\rho = \rho(x, \eta, t)$ and employed the subscript notation $\mathcal{E}_\hbar(t) = \mathcal{E}_\hbar(\infty)\left(1 - e^{-\frac{t}{t_c}}\right)$ to display the quantum nature of the diffusion energy which turns out to be now expressed in terms of the Planck constant ħ. Notice that both the constants $k$ and $\gamma$ as well as the dimensionless function $I(t) = 1 - e^{-\frac{t}{t_c}}$ hold classical during the classical-to-quantum transition.

By making use of the Fourier transform

$$W(x,p,t) = \frac{1}{2\pi}\int_{-\infty}^{\infty}\rho(x,\eta,t)e^{ip\eta}d\eta, \tag{38}$$

which changes the variables from quantum configuration space $(x, \eta\hbar)$ onto quantum phase space $(x, p; \hbar)$, equation of motion (37) turns out to be written down as

$$\frac{\partial W}{\partial t} = \frac{k}{2m\gamma}x\frac{\partial W}{\partial x} - \frac{k}{2m\gamma}p\frac{\partial W}{\partial p} + \frac{\mathcal{E}_\hbar(t)}{2m\gamma}\frac{\partial^2 W}{\partial x^2} + \left[\frac{k}{2m\gamma} - \frac{2\mathcal{E}_\hbar(t)}{m\gamma\hbar^2}p^2\right]W \tag{39}$$

in terms of the phase-space function $W \equiv W(x,p,t)$. Because the exponential factor $e^{ip\eta}$ in (38) is to be a dimensionless term, it follows that the variable $p$ is to have dimensions of linear momentum.

### *4.1.1. The initial condition*

Our quantum phase-space Smoluchowski equation (39) may be solved starting from the non-thermal initial condition

$$W(x,p,t=0) = \frac{1}{\pi\hbar}e^{-\left(\frac{ap^2}{\hbar}+\frac{x^2}{\hbar a}\right)} \tag{40}$$

leading to the distribution $F(x,p) = \delta(x)\delta(p)$ in the classical limit, $\hbar \to 0$. The constant $a$ has dimensions of time per mass, that is, we may set $a = 1/m\gamma$, for instance. The Gaussian probability distribution function (40) gives rise to both variances $\langle P^2(0)\rangle = \frac{\hbar m\gamma}{2}$ and $\langle X^2(0)\rangle = \frac{\hbar}{2m\gamma}$, which fulfill the Heisenberg fluctuation principle $\sqrt{\langle P^2(0)\rangle\langle X^2(0)\rangle} = \frac{\hbar}{2}$.

### *4.1.2. The time-dependent solution*

A time-dependent solution to (39) reads

$$W(x,p,t) = \frac{1}{\pi\hbar}e^{-\frac{p^2}{\hbar^2}B_\hbar(t)}e^{-\frac{x^2}{B_\hbar(t)}} \tag{41}$$

with

$$B_{\hbar}(t) = \frac{\hbar}{m\gamma} e^{-\frac{t}{t_r}} + \frac{2\mathcal{E}_{\hbar}(\infty)}{k} \left\{ 1 - e^{-\frac{t}{t_r}} + \frac{t_c}{t_c - t_r} \left( e^{-\frac{t}{t_r}} - e^{-\frac{t}{t_c}} \right) \right\} \tag{42}$$

expressed in terms of the evolution time $t$, the relaxation time $t_r = m\gamma/k$, as well as the correlation time $t_c$ .

Solution (41) yields $\langle P(t)\rangle = \langle X(t)\rangle = 0$ and the following fluctuations

$$\Delta P(t) = \sqrt{\langle P^2(t)\rangle - \langle P(t)\rangle^2} = \frac{\hbar}{\sqrt{2B_{\hbar}(t)}}, \tag{43}$$

$$\Delta X(t) = \sqrt{\langle X^2(t)\rangle - \langle X(t)\rangle^2} = \sqrt{\frac{B_{\hbar}(t)}{2}}, \tag{44}$$

satisfying the Heisenberg constraint $\Delta P(t)\Delta X(t) = \frac{\hbar}{2}$. The fluctuation (43) gives rise to the concept of thermal force

$$\mathcal{F}(t) = \frac{d\Delta P(t)}{dt} = -\frac{\hbar}{[2B_{\hbar}(t)]^{\frac{3}{2}}} \frac{dB_{\hbar}(t)}{dt} \tag{45}$$

whereas (44) generates the thermal velocity

$$\mathcal{V}(t) = \frac{d\Delta X(t)}{dt} = \frac{1}{2[2B_{\hbar}(t)]^{\frac{1}{2}}} \frac{dB_{\hbar}(t)}{dt}, \tag{46}$$

where

$$\frac{dB_{\hbar}(t)}{dt} = -\frac{k\hbar}{m^2\gamma^2} e^{-\frac{t}{t_r}} + \frac{2\mathcal{E}_{\hbar}(\infty)}{m\gamma} \frac{t_c}{t_r - t_c} \left( e^{-\frac{t}{t_c}} - e^{-\frac{t}{t_r}} \right). \tag{47}$$

Since $B_{\hbar}(t=0) = \frac{\hbar}{m\gamma}$ and $\left.\frac{dB_{\hbar}(t)}{dt}\right|_{t=0} = -\frac{k\hbar}{m^2\gamma^2}$, both quantities (45) and (46) are well defined at $t = 0$, i.e., $\mathcal{F}(t=0) = \frac{k}{2}\sqrt{\frac{\hbar}{2m\gamma}}$ and $\mathcal{V}(t=0) = -\frac{k}{2m\gamma}\sqrt{\frac{\hbar}{2m\gamma}}$. This upshot implies that both fluctuations (43) and (44) are differentiable functions at $t = 0$.

*4.1.3. The long-time regime*

In the long-time regime $t \to \infty$, physically interpreted as $t \gg t_c, t_r$, the probability distribution function (41) renders steady, i.e,

$$W_{\mathrm{st}}(x,p) = \frac{1}{\pi\hbar} e^{-\frac{2\mathcal{E}_{\hbar}(\infty)p^2}{\hbar^2}} e^{-\frac{x^2}{2\mathcal{E}_{\hbar}(\infty)}}, \tag{48}$$

while the fluctuations (43) and (44) become, respectively, $\Delta P(\infty) = \frac{\hbar}{2}\sqrt{\frac{k}{\mathcal{E}_\hbar(\infty)}}$ and $\Delta X(\infty) = \sqrt{\frac{\mathcal{E}_\hbar(\infty)}{k}}$. In contrast, we find out $\mathcal{F}(\infty) = \mathcal{V}(\infty) = 0$, meaning that both quantities (45) and (46) are non-equilibrium effects.

## *4.2. Thermal systems*

### *4.2.1. Quantum limit*

We now assume that our Brownian harmonic oscillator is immersed in a heat bath consisting of quantum harmonic oscillators having the same oscillation frequency $\omega$, so that the Brownian particle's quantum diffusion energy $\mathcal{E}_\hbar(\infty)$ can be identified with the heat bath's quantum thermal energy $\overline{E} = \frac{\omega\hbar}{2}\coth\left(\frac{\omega\hbar}{2k_B T}\right)$, i.e., $\mathcal{E}_\hbar(\infty) = \frac{\omega\hbar}{2}\coth\left(\frac{\omega\hbar}{2k_B T}\right)$ [see Appendix A]. In the quantum limit, i.e., at zero temperature, the thermal diffusion energy $\mathcal{E}_\hbar(\infty)$ should be replaced with $\mathcal{E}_\hbar(\infty) = \frac{\omega\hbar}{2}$ in the formulae above.

### *4.2.2. Classical limit*

In the classical limit $\hbar \to 0$, the time-dependent solution (41) becomes $\mathcal{F}(x,p,t) \propto \delta(p)e^{-\frac{x^2}{B(t)}}$ that in turn leads to the function

$$f(x,t) = \int_{-\infty}^{\infty} \mathcal{F}(x,p,t)dp = \frac{1}{\sqrt{\pi B(t)}} e^{-\frac{x^2}{B(t)}}, \tag{49}$$

which is a solution of the non-Markovian Smoluchowski equation (12) for classical thermal systems. In (49), $B(t)$ is the function

$$B(t) = \frac{2k_B T}{k}\left[1 - e^{-\frac{t}{t_r}} + \frac{t_c}{t_c - t_r}\left(e^{-\frac{t}{t_r}} - e^{-\frac{t}{t_c}}\right)\right], \tag{50}$$

obtained from (42) as $\lim_{\hbar\to 0} B_\hbar(t) = B(t)$.

In the classical limit the root mean square momentum (43) vanishes, whereas (46) becomes

$$\Delta X(t) = \sqrt{\frac{k_B T}{k}\left[1 - e^{-\frac{t}{t_r}} + \frac{t_c}{t_c - t_r}\left(e^{-\frac{t}{t_r}} - e^{-\frac{t}{t_c}}\right)\right]}, \tag{51}$$

breaking down the Heisenberg constraint, i.e., $\Delta P(t)\Delta X(t) = 0$, as expected. This implies that in the classical realm the momentum becomes a deterministic variable while the displacement holds as a stochastic one. In other words, the momentum variable is said to be eliminated from the classical description of Brownian motion in the absence of inertial force.

It is worth stressing that the root mean square displacement (51) is differentiable at short times, i.e., $\mathcal{V}(t) \sim \sqrt{\frac{k_B T}{2m\gamma t_c}}$. Only in the Markovian limit $t_c \to 0$, (51) renders nondifferentiable.

Furthermore, we can derive the Maxwell—Boltzmann distribution function $f(x) = \frac{1}{\sqrt{2\pi k_B T}} e^{-\frac{kx^2}{2k_B T}}$ for the thermal position $x = x' - \frac{2}{k}\sqrt{2m\gamma k_B T}\langle\Psi(t)\rangle$ as the classical limit of the (marginal) probability distribution function (48).

## 5. Application: quantum tunneling

On account of the environmental fluctuations, the escape rate of a Brownian particle over a barrier separating two metastable states— in a double-well potential, for instance— is known as the Kramers problem [17,21,26]. In the classical domain, the transition of such an particle over the potential barrier from a metastable state toward another state is said to be an activation phenomenon, while the corresponding metastability phenomenon in the quantum domain characterizes a tunneling process. In both realms of physics, escape rate phenomena are non-equilibrium effects taking place in the (Markovian) steady regime.

In 1940 Kramers [21] extended the Einstein—Langevin—Kolmogorov approach to study the issue of metastability of a Brownian particle in phase space and found an escape rate characterized by non-quantum, thermal, and steady effects. In [27] we have generalized Kramers' theory to open classical systems far from thermal equilibrium. In this section, we wish to extend Kramers' technique to our quantum Smoluhowski equation (39) reckoning with two physical situations: Section 5.1 deals with non-thermal systems and section 5.2 considers thermal systems. Section 5.3 brings some discussions.

### *5.1. Non-thermal open quantum systems*

In order to calculate the quantum Kramers escape rate of a non-inertial Brownian particle over a potential barrier, we consider as a starting point the steady solution (48) in the form

$$W_{\mathrm{st}}(x,p) = \frac{1}{\pi\hbar} e^{-\left[\frac{V(x)}{\mathcal{E}_\hbar(\infty)} + \frac{2\mathcal{E}_\hbar(\infty)p^2}{k\hbar^2}\right]}. \qquad (52)$$

We perform an expansion of the potential function $V(x) = kx^2/2$ in a Taylor series around a certain point $x_0$, i.e., $V(x) \sim V(x_0) + (k/2)(x - x_0)^2$, and then construct the steady non-equilibrium function

$$W_{\text{neq}}(x,p) = \alpha\varphi(x,p)e^{-\frac{\left[V(x_0)+\frac{k}{2}(x-x_0)^2\right]}{\mathcal{E}_\hbar(\infty)} - \frac{2\mathcal{E}_\hbar(\infty)}{\hbar^2 k}p^2}, \tag{53}$$

$\alpha$ being a constant and $\varphi(x,p)$ a function to be determined. On inserting the function (53) into our quantum Smoluchowski equation (39), given by

$$\frac{k(x-x_0)}{2}\frac{\partial W_{\text{neq}}}{\partial x} - \frac{kp}{2}\frac{\partial W_{\text{neq}}}{\partial p} + \frac{\mathcal{E}_\hbar(\infty)}{2}\frac{\partial^2 W_{\text{neq}}}{\partial x^2} + \left[\frac{k}{2} - \frac{2\mathcal{E}_\hbar(\infty)}{\hbar^2}p^2\right]W_{\text{neq}} = 0, \tag{54}$$

with $W_{\text{neq}} \equiv W_{\text{neq}}(x,p)$, we obtain the equation of motion

$$\mathcal{E}_\hbar(\infty)\frac{\partial^2\varphi(x,p)}{\partial x^2} - k(x-x_0)\frac{\partial\varphi(x,p)}{\partial x} - kp\frac{\partial\varphi(x,p)}{\partial p} = 0, \tag{55}$$

which changes into

$$\frac{d^2\varphi(\xi\,)}{d\xi^2} = \frac{k}{\beta^2\mathcal{E}_\hbar(\infty)}\xi\frac{d\varphi(\xi\,)}{d\xi} \tag{56}$$

in terms of the new variable $\xi$ given by

$$\xi = \beta(x - x_0) - p. \tag{57}$$

The constant $\beta$ in (56) and (57) is to have dimensions of mass per time. Hence, we may put $\beta = m\gamma/2$, for instance. A solution to the differential equation (56) reads

$$\varphi(\xi) = \frac{2}{m\gamma}\sqrt{\frac{-k}{2\pi\mathcal{E}_\hbar(\infty)}}\int_{-\infty}^{\xi} e^{\frac{2k\xi^2}{m^2\gamma^2\mathcal{E}_\hbar(\infty)}}d\xi\ , \tag{58}$$

fulfilling the condition $\varphi(\xi \to \infty) = 1$. This result (58) is only possible if the potential curvature is negative, i.e., $k < 0$. Hence, we hereafter use $x_0 \equiv x_b$, and $k = -m\omega_b^2$, where the quantity $\omega_b$ denotes the particle's oscillation frequency over the potential barrier at $x_b$. So the barrier top is located at point $x_b$, while the two bottom wells are at $x_a$ and $x_c$, such that $V(x_c\,) = 0 = V(x_a)$, with $x_a\ < x_b$.

Using (58), the stationary function (53) becomes

$$W_{\text{neq}}(x,p) = \frac{1}{\gamma}\frac{2\alpha\omega_b}{\sqrt{2\pi m\mathcal{E}_\hbar(\infty)}}e^{-\frac{\left[V(x_b)-\frac{m\omega_b^2}{2}(x-x_b)^2\right]}{\mathcal{E}_\hbar(\infty)} + \frac{2\mathcal{E}_\hbar(\infty)}{\hbar^2 m\omega_b^2}p^2}\int_{-\infty}^{\xi} e^{\frac{2k\xi^2}{m^2\gamma^2\mathcal{E}_\hbar(\infty)}}d\xi \quad (59$$

from which we can find out the probability current as

$$J_b = \frac{1}{m}\int_{-\infty}^{\infty} pW_{\text{neq}}(x = x_b, p)\,dp = -\frac{\alpha\omega_b^4\hbar^3}{4\mathcal{E}_\hbar(\infty)}\frac{e^{-\frac{V(x_b)}{\mathcal{E}_\hbar(\infty)}}}{\sqrt{\hbar^2\omega_b^4 - \gamma^2\mathcal{E}_\hbar^2(\infty)}}, \tag{60}$$

where we used the result

$$\int_{-\infty}^{\infty} e^{-up^2}\,pdp \int_{-\infty}^{\xi=p} e^{-v\xi^2}d\xi = \frac{1}{2u}\left(\frac{\pi}{u+v}\right)^{1/2},$$

$u$ and $v$ being non-negative constants.

The number of particles through the well located at $x_a$ is calculated as $n_a = \iint_{-\infty}^{\infty} W_{\text{eq}}(x,p)dpdx = \alpha\pi\hbar$, whereby we used the equilibrium distribution function $W_{\text{eq}}(x,p) = \alpha e^{-\frac{4\mathcal{E}_\hbar(\infty)p^2}{\hbar^2 m\omega_a^2} - \frac{m\omega_a^2}{4\mathcal{E}_\hbar(\infty)}(x-x_a)^2}$ coming from the non-equilibrum function (53) for $\varphi(x,p) = 1$ calculated at the well $x_a$, with $k_a = m\omega_a^2$.

Making use of (60) and $n_a = \alpha\pi\hbar$, the quantum Kramers escape rate of a Brownian particle immersed in a non-thermal environment is then written down as

$$\Gamma(\infty) = \frac{|J_b|}{n_a} = \frac{\omega_b^4\hbar^2}{4\pi\mathcal{E}_\hbar(\infty)}\frac{e^{-\frac{V(x_b)}{\mathcal{E}_\hbar(\infty)}}}{\sqrt{\hbar^2\omega_b^4 - \gamma^2\mathcal{E}_\hbar^2(\infty)}} \tag{61}$$

fulfilling the condition $0 < \gamma\mathcal{E}_\hbar(\infty) < \hbar\omega_b^2$.

It is worth highlighting that our main upshot (61) does display dimensions of inverse of time, as it should be expected, as well as being compatible with the quantum dissipation-fluctuation relation $\mathcal{E}_\hbar(\infty)$.

Our quantum escape rate (61) decays exponentially with respect to the ratio$-V(x_b)/\mathcal{E}_\hbar(\infty)$. In addition, the so-called quantum Arrhenius factor is given by $\sigma = \frac{1}{4\pi\mathcal{E}_\hbar(\infty)}\frac{\omega_b^4\hbar^2}{\sqrt{\hbar^2\omega_b^4-\gamma^2\mathcal{E}_\hbar^2(\infty)}}$.

*5.2. Thermal open quantum systems*

For thermal systems, i.e., $\mathcal{E}_\hbar(\infty) = \frac{\omega_b\hbar}{2}\coth\left(\frac{\omega_b\hbar}{2k_BT}\right)$, our quantum Kramers rate (61) becomes

$$\Gamma(\infty) = \frac{\omega_b^2}{\pi\coth\left(\frac{\omega_b\hbar}{2k_BT}\right)}\frac{e^{-\frac{2V(x_b)}{\omega_b\hbar\coth\left(\frac{\omega_b\hbar}{2k_BT}\right)}}}{\sqrt{4\omega_b^2 - \gamma^2\coth^2\left(\frac{\omega_b\hbar}{2k_BT}\right)}} \tag{62}$$

which is valid for low temperatures: $\gamma \coth\left(\frac{\omega_b \hbar}{2k_B T}\right) < 2\omega_b$. At zero temperature equation (62) reads

$$\Gamma^{(T=0)}(\infty) = \frac{\omega_b^2}{\pi} \frac{e^{-\frac{2V(x_b)}{\omega_b \hbar}}}{\sqrt{4\omega_b^2 - \gamma^2}}, \tag{63}$$

with $\gamma < 2\omega_b$. Both results (62) and (63) show how dissipative effects affect the quantum tunneling process in the Smoluchowski regime. This dissipative tunneling phenomenon is in accordance with Ankerhold's complaints [28]. It is worth noticing that dissipative effects in both (62) and (63) account for enhancing the quantum tunneling rate. This outcome is in contrast to the classical Kramers escape rate [21]

$$\Gamma(\infty) = \frac{\omega_a \omega_b}{4\pi\gamma} e^{-\frac{V(x_b)}{k_B T}}$$

in which damping effects are responsible for diminishing the Arrhenius factor.

*5.3. Discussions*

Because of the wide practical and theoretical interest in chemical physics, for instance, recent studies have been come out showing how quantum effects upon the Smoluchowski equation and the Kramers rate can arise.

First, on the basis of the Caldeira—Leggett Hamiltonian system-plus-reservoir model [5,8,10] in tandem with the path integral formalism, Ankerhold and co-workers [8,29] have derived a quantum Smoluchowski equation (QSE) and explored its physical meaning applying it to chemical reactions, mesoscopic physics, and charge transfer in molecules, for instance. Meanwhile, further works [30] have revealed some drawbacks underlying such a QSE, for it may violate the Second Law of Thermodynamics, for example. It is relevant to point out that Ankerhold's QSE for the case of a free Brownian particle coincides with the classical Smoluchowski equation .

Another Hamiltonian system-plus-reservoir account aiming at to quantize the Smoluchowski equation has been developed in [31]. Here a QSE, which for free Brownian motion differs from the classical Smoluchowski equation, is derived and the problem of quantum tunneling is studied at zero temperature, in contrast to Ankerhold's survey. This Indian group's approach does not depend on the path integral formalism but is based on the canonical quantization.

Coffey and co-workers [32] in turn have obtained a QSE in the semi-classical region of quantum mechanics on the basis of Brinkman's hierarchy in terms of the time evolution for the Wigner function obtained from the Caldeira—Leggett Hamiltonian system-plus-reservoir model. Although Coffey's QSE can be applied to the dynamics of a point Josephson junction, for instance, it is plagued with the same disease as Ankerhold's QSE, namely, both quantum and classical dynamics for free Brownian motion are identical.

Starting from the quantum linear Boltzmann equation, Vacchini [33] has derived a QSE in which the dynamics of a free quantum particle is distinct from a classical free particle.

Lastly, Tsekov [34] has arrived at a version of QSE without referring to any quantization process. He starts with a Hamiltonian system-plus-reservoir model described by a non-linear Schrödinger equation in the Madelung representation of quantum mechanics and then obtains a QSE for a free Brownian particle different from the classical Smoluchowski equation, too.

Alternatively, in previous works [5,14] and in the present study we have derived a quantum Smoluchowski equation based on a non-Hamiltonian quantization method without making use of the canonical quantization as well as the path integral formalism. Our QSE (39) for a free particle ($k = 0$) does not coincides with the classical Smoluchowski equation. Moreover, our equation (39) can be applied to looking at quantum tunneling process at zero temperature (see result (63)).

## 6. Summary and outlook

In this paper we have taken up the dynamical-quantization approach to quantum open systems without assuming the existence of a Hamiltonian framework underlying the interaction between a Brownian particle and its environment. Our non-Hamiltonian approach features the following upshots:

(a) By quantizing the non-Gaussian Kolmogorov equation in phase space (4), we have derived the quantum master equation (15) describing the quantum Brownian motion of a particle immersed in a generic environment (e.g., a non-thermal quantum fluid). For the case of a thermal quantum reservoir, our master equation (15) in the Gaussian approximation yields the non-Markovian Caldeira—Leggett equation (17) that has been solved for a particle in a gravitational field. Here, we have put forward the concept of thermal quantum force (28) that is a non-equilibrium effect carrying both non-Markovian and relaxation features. Moreover, we have shown that Markovian effects account for the differentiability property of the root mean square momentum (26) in the classical limit.

(b) Quantizing the non-Gaussian Kolmogorov equation in configuration space (11) leads to the quantum equation of motion (C.4) reducing to (C.6)in the Gaussian approximation (see Appendix C). For the case of a harmonic oscillator immersed in a general non-thermal environment, our Gaussian quantum equation (C.6) gives rise to the non-Markovian quantum Smoluchowski equation in phase space (39) that has been solved. For a thermal quantum heat bath we have derived the thermal quantum force (45) as well as the thermal quantum velocity (46) that are well defined at $t \geq 0$ and at zero temperature. In the classical limit (in the high-temperature regime) the root mean square momentum (43) vanishes meaning that the momentum variable is a deterministic one in the non-inertial classical Brownian motion. Moreover, the classical limit of (43) yields the quantity (51) that is differentiable for all time owing to non-Markovian effects. By contrast, in the Markovian limit the root mean square displacement (51) renders non-differentiable, as it is well known since Einstein's 1905 paper.

(c) Our quantum Smoluchowski equation (39) has been also applied to describing the tunneling process of a non-inertial quantum Brownian particle in a non-thermal dissipative medium. Our main finding is the quantum Kramers rate (61). For thermal reservoir (61) becomes (62) that in turn is valid in the low-temperature regime (including the zero temperature case). According to our quantum escape rates (62) and (63) dissipative effects account for enhancing the quantum tunneling rate.

Lastly, in a forthcoming article we intend to tackle the problem of quantum Brownian motion in the presence of bosonic and fermionic heat baths as well as non-Gaussian environments as described by our quantum Kolmogorov equations (15) and (C.4).

## Acknowledgments

I would like to thank Professor Maria Carolina Nemes (Universidade Federal de Minas Gerais, Brazil) for the scientific support as well as the Fundação de Amparo à Pesquisa do Estado de Minas Gerais (Fapemig, Contract CEX-00103/10) for the financial support. Referee's constructive criticisms will be acknowledged, too.

## Appendix A. The physis of the environment

Imagine that the environment is made up of a set of quantum harmonic oscillators having the same oscillation frequency $\omega$. It is readily to show that the mean energy of this quantum heat bath after attaining the thermodynamic equilibrium at temperature $T$ is given by [35]

$$\overline{E} = \frac{\omega\hbar}{2}\left(\frac{e^{\frac{\omega\hbar}{k_B T}}+1}{e^{\frac{\omega\hbar}{k_B T}}-1}\right) = \frac{\omega\hbar}{2}\coth\left(\frac{\omega\hbar}{2k_B T}\right),$$

where $k_B$ is dubbed Boltzmann's constant. The ħ-dependent energy, $\omega\hbar/2$, corresponds to the zero point energy of the heat bath at zero temperature and $k_B T$ denotes the classical thermal energy of the heat bath at high temperature, i.e., $T \gg \omega\hbar/2k_B$.

## Appendix B. The non-Markovian correlational function

The correlational function $I(t)$ is given by

$$I(t) = \lim_{\varepsilon \to 0} \frac{1}{\varepsilon} \iint_{t}^{t+\varepsilon} \langle \Psi(t')\Psi(t'') \rangle dt' dt'' \,.$$

Under the condition $\lim_{t \to \infty} I(t) \approx I(\infty) = 1$, the autocorrelation function $\langle \Psi(t')\Psi(t'') \rangle$ can be built up as

$$\langle \Psi(t')\Psi(t'') \rangle = \left(1 - e^{-\frac{(t\prime + t\prime\prime)}{2t_c}}\right) \delta(t' - t''),$$

where $t_c$ is the correlation time of $\Psi(t)$ at times $t'$ and $t''$. Accordingly, we find the non-Markovian correlational function

$$I(t) = 1 - e^{-\frac{t}{t_c}},$$

reducing to $I(t) = 1$ in the Markovian limit $t_c \to 0$.

## Appendix C. Quantizing the Kolmogorov equation in configuration space

We wish to quantize the non-Gaussian Kolmogorov equation in configuration space (11). To this end, let $\chi_1 = \chi(x_1, t)$ be a solution of the non-Gaussian Kolmogorov equation (11) and $\chi_2 = \chi(x_2, t)$ another one at $x_2$:

$$\frac{\partial\chi(x_1,t)}{\partial t} = \sum_{k=1}^{\infty} \frac{(-1)^k}{k!} \frac{\partial^k}{\partial x_1^k} \left[\overline{A}_k(x_1,t)\chi(x_1,t)\right], \tag{C.1}$$

$$\frac{\partial\chi(x_2,t)}{\partial t} = \sum_{i=1}^{\infty} \frac{(-1)^k}{k!} \frac{\partial^k}{\partial x_2^k} \left[\overline{A}_k(x_2,t)\chi(x_2,t)\right]. \tag{C.2}$$

where $\overline{A}_k(x_1, t)$ and $\overline{A}_k(x_2, t)$ are coefficients associated with solutions $\chi(x_1, t)$ and $\chi(x_2, t)$, respectively. Multiplying (C.1) and (C.2) by $\chi(x_2, t)$ and $\chi(x_1, t)$, respectively, and then adding the resulting equations we arrive at

$$\frac{\partial\xi(x_1,x_2,t)}{\partial t} = \sum_{k=1}^{\infty} \frac{(-1)^k}{k!} \left(\frac{\partial^k}{\partial x_1^k}\overline{A}_k(x_1,t) + \frac{\partial^k}{\partial x_2^k}\overline{A}_k(x_2,t)\right)\xi(x_1,x_2,t) \tag{C.3}$$

where $\xi(x_1, x_2, t) = \chi(x_1, t)\,\chi(x_2, t) = \sqrt{f(x_2, t)f(x_1, t)}$. By quantizing via (14) and making use of the relations $\frac{\partial}{\partial x_1} = \frac{1}{2}\frac{\partial}{\partial x} - \frac{1}{\hbar}\frac{\partial}{\partial \eta}$ and $\frac{\partial}{\partial x_2} = \frac{1}{2}\frac{\partial}{\partial x} + \frac{1}{\hbar}\frac{\partial}{\partial \eta}$, the classical equation (C.3) becomes the quantum Kolmogorov equation in configuration space

$$\frac{\partial\rho(x,\eta,t)}{\partial t} = \mathbb{L}_\hbar\, \rho(x,\eta,t) \tag{C.4}$$

with the ħ-dependent operator $\mathbb{L}_\hbar$ given by

$$\mathbb{L}_\hbar = \sum_{k=1}^{\infty} \frac{(-1)^k}{k!}\left[\left(\frac{\partial}{2\partial x} - \frac{1}{\hbar}\frac{\partial}{\partial\eta}\right)^k A_k^{(\hbar)}\left(x - \frac{\eta\hbar}{2}, t\right) + \left(\frac{\partial}{2\partial x} + \frac{1}{\hbar}\frac{\partial}{\partial\eta}\right)^k A_k^{(\hbar)}\left(x + \frac{\eta\hbar}{2}, t\right)\right] \tag{C.5}$$

The quantum Kolmorov equation (C.4)exhibits non-Gaussian features in full. Yet, in the Gaussian approximation $|x_2 - x_1|^3 \ll 0$, it changes into

$$\frac{\partial\rho}{\partial t} = \frac{1}{m\gamma}\frac{\partial^2\mathcal{V}_{\mathrm{eff}}^{(\hbar)}(x,t)}{\partial x^2}\rho + \frac{1}{2m\gamma}\frac{\partial\mathcal{V}_{\mathrm{eff}}^{(\hbar)}(x,t)}{\partial x}\frac{\partial\rho}{\partial x} + \frac{\eta}{2m\gamma}\frac{\partial^2\mathcal{V}_{\mathrm{eff}}^{(\hbar)}(x,t)}{\partial x^2}\frac{\partial\rho}{\partial\eta}$$
$$+ \frac{\mathcal{E}_\hbar(\infty)}{m\gamma} I(t)\left[\frac{\partial^2\rho}{\partial x^2} + \frac{4}{\hbar^2}\frac{\partial^2\rho}{\partial\eta^2}\right] \tag{C.6}$$

with the effective potential

$$\mathcal{V}_{\mathrm{eff}}^{(\hbar)}(x,t) = V(x) - x\sqrt{4\gamma m \mathcal{E}_{\hbar}(\infty)}\langle \Psi(t)\rangle \tag{C. 7}$$

Considering (C.6) at point $x'$ for a harmonic potential $V(x') = kx'^2/2$, it reduces to equation (37) in terms of the effective position $x = x' - \frac{2}{k}\sqrt{2m\gamma\mathcal{E}_{\hbar}(\infty)}\langle\Psi(t)\rangle$.